\documentclass{iopart}

\usepackage{latexsym}
\usepackage{iopams} 
\usepackage{setstack} 
\usepackage{graphicx}

\newcommand{\bra}[1]{\langle #1 |}
\newcommand{\ket}[1]{| #1 \rangle}

\def\etal{\textit{et al.{}}}
\newcommand{\im}{\dot{\iota}\,}
\begin{document}

\title[Control of atomic entanglement with optical detuned cavity]
      {Control of entanglement and two-qubit quantum gates with
       atoms crossing a detuned optical cavity}

\author{D Gon\c{t}a$^1$ and S Fritzsche$^{2,3}$}

\address{$^1$ Max-Planck-Institut f\"{u}r Kernphysik, \\
              \hspace*{0.28cm} P.O.~Box 103980, D--69029 Heidelberg, Germany}

\address{$^2$ Gesellschaft f{\"u}r Schwerionenforschung (GSI), \\
              \hspace*{0.28cm} D--64291 Darmstadt, Germany}

\address{$^3$ Department of Physical Sciences, \\
              \hspace*{0.28cm} P.O.~Box 3000, Fin-90014 University of Oulu, Finland}

\ead{gonta@physi.uni-heidelberg.de}
\ead{s.fritzsche@gsi.de}

\begin{abstract}
A scheme is proposed to generate an entangled state between two
($\Lambda$-type) four-level atoms that interact effectively by means
of a detuned optical cavity and a laser beam that acts
perpendicularly to the cavity axis. It is shown how the degree of
entanglement for two atoms passing through the cavity can be
controlled by manipulating their velocity and the (initial) distance
between the atoms. In addition, three realistic schemes are
suggested to implement the two-qubit gates within the framework of
the suggested atom-cavity-laser setup, namely, the i-swap gate,
controlled-Z gate, and the controlled-$\overline{\mbox{NOT}}$ gate.
For all these schemes, we analyze and discuss the atomic velocities
and inter-atomic distances for which these gates are realized most
reliably.
\end{abstract}


\maketitle

\section{Introduction}

During recent years, quantum entanglement has been found important
not only in studying the non-classical behavior of composite systems
but also as one essential resource for the engineering and
processing of quantum information. Nowadays, there are  known
various applications that (would) greatly benefit from having
entangled quantum states available as, for instance, super-dense
coding \cite{bew}, quantum cryptography \cite{eke}, or the use of
Grover's quantum search algorithm \cite{gro}, to name just a few of
them. Despite of the recent progress in dealing with composite
quantum systems, however, their manipulation and controlled
interaction with the environment has remained a challenge for
experiment until the present. Apart from various other
implementations of composite systems, the proof for and an excellent
control about the generation of entanglement has been achieved
especially with neutral atoms that are coupled to high-finesse
optical cavities \cite{cp34, ps76, nat424}.

From the experimental perspective, there are two basic types of
(atomic) level configurations utilized to encode and store a single
qubit: Apart from (i) the use of \textit{optical} qubit, that simply
refers to the two atomic levels separated by a optical transition
frequency, one may (ii) utilize also the (so-called)
\textit{hyperfine} qubit that is associated with two hyperfine
levels of--usually--the electronic ground state of the atom. In
neutral atoms, these hyperfine levels are typically separated by a
microwave frequency and are known to be robust with regard to
decoherence effects and external stray fields in contrast to the
optical qubits mentioned above. For the hyperfine qubits, therefore,
rather long coherence times ($\sim$ 1 s) have been reported in the
literature \cite{prl92, prl95, prl91}. In addition, a number of
microwave techniques have been developed during the last decades in
order to initialize, manipulate and detect the state of such
hyperfine qubits \cite{prl92, prl95, prl91, prl93, pra75, pra76}.

Unfortunately, however, a hyperfine qubit cannot couple directly to
a cavity with the resonant mode frequency in the optical domain.
Therefore, in order to manipulate the information encoded by the
atom, the superposition of the two hyperfine levels must first be
transferred coherently to some other two (electronically) excited
states before the atom enters the cavity, and this information must
be brought back in a coherent fashion after the atom exits the
cavity. Instead of an atomic two-level configuration, we then need
to consider a four-level scheme, in which the two hyperfine levels
for storing quantum information are associated with two (additional)
optically excited levels. In order to realize an efficient
atom-cavity coupling, moreover, the energy splitting of the two
electronically excited levels should be compatible with the resonant
frequency of the cavity. In this way, a coupling between the
hyperfine qubit and the optical cavity can be achieved and might
open a route towards the implementation of quantum gates via
cavity-mediated atom-atom interactions.

The basic idea for the formation of entanglement between two
four-level atoms that are coupled to a optical cavity and a external
laser beam have been first suggested by You and co-authors
\cite{pra67}, which follows the novel cavity-mediated atom-atom
interaction regime from Ref. \cite{prl85}. This interaction regime
is based on the exchange of a photon between two bi-level atoms that
is stimulated by a cavity which is detuned with regard to transition
frequencies of the atoms. Later in Ref.~\cite{prl87}, moreover, it
was shown experimentally how this effective atom-atom interaction
leads to the generation of entanglement between two atoms that cross
a detuned microwave cavity. In the theoretical analysis of You
\etal{}, however, it was assumed that both atoms couple to the same
cavity mode via a constant coupling strength being independent on
the atomic position inside the cavity mode. In practice, the atoms
cross the cavity one after another being separated by a macroscopic
distance. This separation of the atoms implies that they have
different atom-cavity couplings as given by the radiation pattern of
the cavity mode standing-wave. Therefore, a more detailed
description of the cavity-mediated entanglement formation has to be
considered, in which the degree of entanglement between the atoms
depends also on the atom-cavity coupling which depends, in turn, on
the location of both atoms inside the cavity mode. Such a
position-dependent coupling between the atoms and the cavity
requires a revision of the previous theoretical analysis and
suggests that the degree of entanglement, that is finally obtained,
might depend substantially on the details of how the atoms cross the
cavity in the course of interaction.

In the present work, we propose a scheme to generate an entangled
state between the hyperfine qubits of two four-level atoms in a
$\Lambda$-type level configuration. In this scheme, the interaction
between the atoms is mediated by a optical cavity and a laser beam
that acts perpendicularly to the cavity axis. In contrast to the
analysis made by You \etal{}, moreover, we assume (i) the atoms to
be separated from each other by a macroscopic distance such that
\textit{no direct} interaction between the atoms occurs and that
(ii) both atoms interact simultaneously with the same cavity mode
and laser field via a position-dependent couplings while passing
through the cavity-laser setup. This scheme leads to an cavity-laser
mediated \textit{effective} atom-atom interaction, in which the two
parameters to control this interaction are the velocity that encodes
the atom-cavity interaction time, and the distance between the atoms
that encodes the atom-cavity interaction strength. For two atoms
being initially prepared in a product state (of their hyperfine
qubits), we determine those velocities and distances for which the
atoms become maximally entangled when passing through the proposed
setup. Apart from generating entanglement between the atoms, we also
suggest schemes to implement various two-qubit quantum logical
gates, such as the i-swap gate, controlled-Z gate, and
controlled-$\overline{\mbox{NOT}}$ gate. All these gates can be
implemented within the given framework, and the respective velocity
and inter-atomic distance of the atoms are discussed, for which the
gate fidelities become maximal.

The paper is organized as follows. In the next section, we introduce
the scheme to entangle the hyperfine qubits of two four-level atoms.
This includes the theoretical description of the effective atom-atom
interaction evolution that allows to control this interaction in
practice. In Sec.~2.1 , in particular, we present and explain all
the steps necessary within the proposed (experimental) set-up, while
a more detailed view on this effective interaction is given in
Sec.~2.2 by using the adiabatic elimination procedure. In Sec.~3,
then, the schemes for the implementation of the i-swap,
controlled-Z, and controlled-$\overline{\mbox{NOT}}$ gates are
presented and discussed. A few conclusions are finally given in
Sec.~4.

\begin{figure}
\begin{center}
\includegraphics[width=0.75\textwidth]{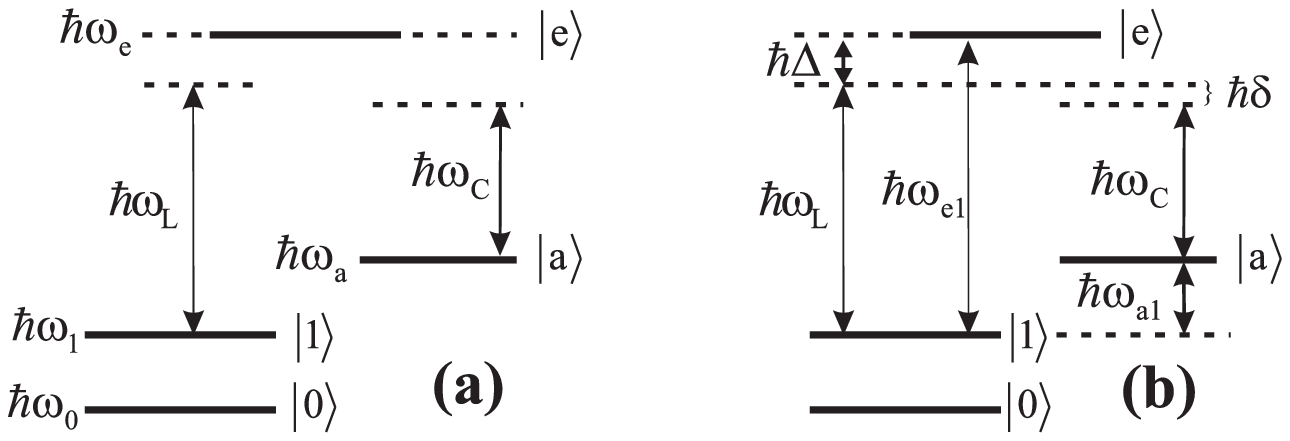} \\
\vspace{0.25cm}
\includegraphics[width=0.75\textwidth]{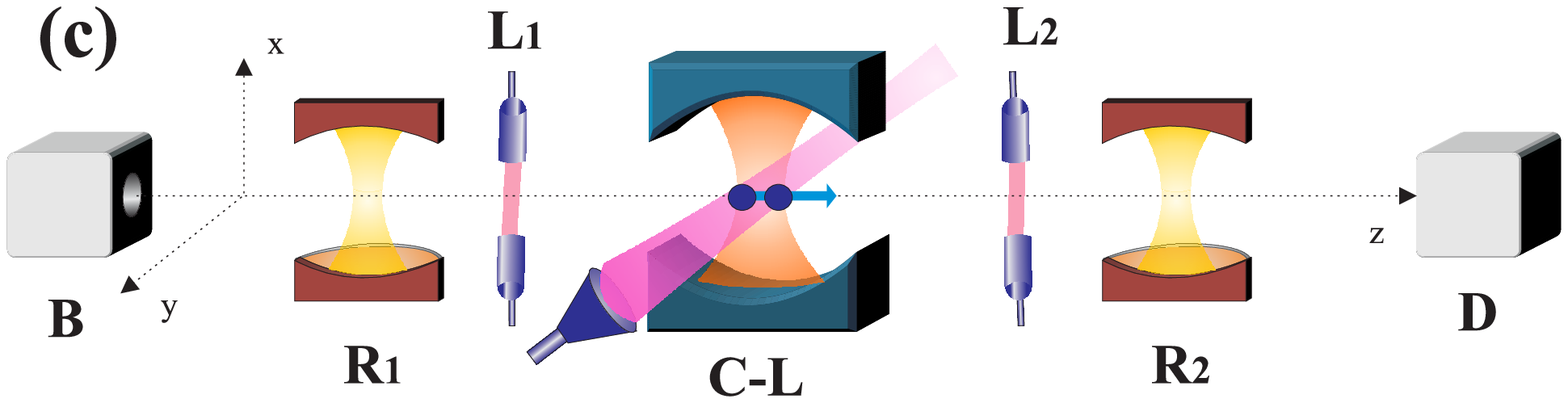} \\
\caption{(Color online) The atomic four-level $\Lambda$-type
configuration in (a) the Schr\"{o}dinger picture and (b) the
interaction picture. (c) Schematic setup of the experiment. Two
neutral atoms from a source $B$ are supposed to pass through a
Ramsey zone $R_1$, a pair of Raman laser beams $L_1$, an optical
cavity $C$ with a perpendicularly acting laser beam $L$ as well as
through a second pair of Raman lasers $L_2$ and Ramsey zone $R_2$,
before the hyperfine states of the atoms is detected at the detector
$D$.}
\label{fig:1}
\end{center}
\end{figure}

\section{Generation of the Two-Atom Entanglement via Optical Cavity}

In this section, we propose and explain our scheme to entangle the
hyperfine qubits of two four-level atoms if they were initially
prepared in a product state. We hereby assume that the atoms can be
controlled with regard to their separation and velocity when they
enter the experimental setup that is displayed in
Fig.~\ref{fig:1}(c).

\subsection{Off-Resonant Atom-Cavity interaction}

\begin{figure}
\begin{center}
\includegraphics[width=0.55\textwidth]{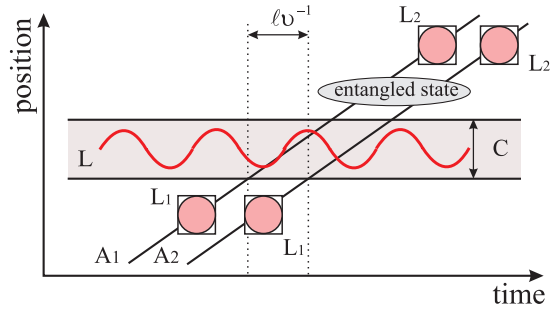} \\
\caption{(Color online) Temporal sequence of steps that need to be
carried out in order to generate an entangled state for the two
hyperfine qubits of atoms $A_1$ and $A_2$. The grey rectangular area
$C$ denotes the spatial extent of the cavity. The (pink) boxed
circles, denoted as $L_1$ and $L_2$, refer to the two (pairs of)
Raman laser beams in front and behind the cavity. The two atomic
qubits are entangled with each other when both atoms have left the
cavity.}
\label{fig:2}
\end{center}
\end{figure}

Let us start by considering an atom in the $\Lambda$-type four-level
configuration as displayed in Fig.~\ref{fig:1}(a). In this level
configuration, the two (hyperfine) states $\ket{0}$ and $\ket{1}$ of
the atomic ground levels carry the qubit information and are
supplemented by the two electronically excited states $\ket{a}$ and
$\ket{e}$ that are separated from each other by an optical
transition frequency. Below, we assume to have two identical atoms
$A_1$ and $A_2$ in such a $\Lambda$-type configuration that are
initially prepared in the composite state $\ket{0,\bar{1}} \equiv
\ket{0} \times \ket{\bar{1}}$, where the bar in $\ket{\bar{\alpha}}$
refers to the state of atom $A_2$. In addition, the atoms are
separated by a macroscopic distance $\ell$ being large enough such
that they do not interact directly with each other and both atoms
move with the same (constant) velocity $\vec{\upsilon}$ along the
$z$ axis [see Fig.~\ref{fig:1}(c)]. Before atom $A_1$ enters the
cavity, its electronic population is transferred from state
$\ket{0}$ to the state $\ket{a}$ by using a pair of slightly
off-resonant laser beams which are coupled to the atomic transitions
$\ket{0} \leftrightarrow \ket{e}$ and $\ket{e} \leftrightarrow
\ket{a}$, respectively. Such a population transfer is known as the
two-photon Raman process \cite{pr155} that enables one to perform a
second-order transition between the states $\ket{0}$ and $\ket{a}$.
For instance, this could be done by utilizing a two phase-locked
laser diode \cite{pra75}. Below, we shall refer to this population
transfer briefly as a Raman pulse and will distinguish between the
Raman pulses (laser beams) $L_1$ and $L_2$ in front and behind the
cavity [see Fig.~\ref{fig:1}(c)]; also in the temporal diagram from
Fig.~\ref{fig:2}, these Raman pulses are displayed as boxed pink
circles. We assume, therefore, that the purpose of these Raman
pulses is just to transfer the electronic population from hyperfine
state $\ket{0}$ to the optical level $\ket{a}$ in the zone $L_1$,
and back from $\ket{a}$ to $\ket{0}$ in zone $L_2$. The same Raman
pulses are applied also to the atom $A_2$ which follows $A_1$
subsequently with distance $\ell$. However, since the second atom
enters the set-up in the state $\ket{\bar{1}}$, it remains
unaffected by the Raman pulse $L_1$ and, thus, the two atoms enter
the cavity in the product state $\ket{a,\bar{1}}$.

Inside the cavity, both atoms $A_1$ and $A_2$ couple via the optical
transition $\ket{a} \leftrightarrow \ket{e}$ (and $\ket{\bar{a}}
\leftrightarrow \ket{\bar{e}}$, respectively) to the same cavity
mode with the resonant frequency $\omega_C$ [see
Fig.~\ref{fig:1}(a)]. As we discussed above, a revised description
of the atom-cavity interaction evolution is based on the
position-dependent atom-cavity coupling
\begin{equation}\label{coupling}
g(\vec{r}) = g_o \, \exp \left( - |\vec{r}|^2 / w^2 \right),
\end{equation}
where $g_o$ denotes the vacuum Rabi frequency and $w$ the
(so-called) cavity mode waist that is the minimum width of the
radiation pattern given by the cavity mode standing-wave. For the
two atoms which move through the cavity with the velocity
$\vec{\upsilon}$ along the $z$ axis, the Gaussian profile
(\ref{coupling}) gives rise to the time-dependent atom-cavity
couplings $g_1(t) \equiv g(z_1^o + \upsilon \, t)$ and $g_2(t)
\equiv g(z_2^o + \upsilon \, t)$, and where $z^o_1 - z^o_2 = \ell >
0$ denotes the initial distance between the atoms.

The cavity-mediated atom-atom interaction (i.e., without the laser
beam $L$) is based on the stimulated exchange of a \textit{single}
photon between two atoms prepared in the product state $\ket{a,
\bar{e}}$. This photon exchange can be understood as the emission of
a virtual photon into the cavity mode by the atom $A_2$ and the
re-absorbtion of the photon by the atom $A_1$, while both atoms are
coupled off-resonantly to the same cavity mode. An off-resonant
atom-cavity interaction hereby refers to the case when the
difference (or detuning) between the atomic $\ket{a} \leftrightarrow
\ket{e}$ transition frequency and the frequency of the cavity mode
$\omega_L$ is large enough: $|\omega_C - (\omega_e - \omega_a)| \gg
|g_\mu(t)|$, so that only a virtual atom-cavity energy exchange can
occur \cite{prl85}.

In our present scheme, in contrast, the atoms enter the cavity in
the composite state $\ket{a, \bar{1}}$, and hence a further
intermediate process $\ket{a,\bar{1}} \rightarrow  \ket{a, \bar{e}}$
is first necessary to obtain the state $\ket{a, \bar{e}}$ that could
evolve into $\ket{e, \bar{a}}$ by means of the detuned cavity. For
this reason, the atoms are exposed to a laser beam that acts
transversally to the cavity axis and \textit{in addition} to their
interaction with the cavity mode [see Fig.~\ref{fig:1}(c)]. The
laser given by the frequency $\omega_L$ couples the atomic
transitions $\ket{1} \leftrightarrow \ket{e}$ and $\ket{\bar{1}}
\leftrightarrow \ket{\bar{e}}$, respectively, as shown in
Fig.~\ref{fig:1}(a). The position-dependent atom-laser coupling
$\Omega(\vec{r}) \,=\, \Omega_o \, \exp \left( - |\vec{r}|^2 /
 \widetilde{w}^2 \right)$ hereby implies the time-dependent
couplings for each atom, namely $\Omega_1(t) \equiv \Omega(z_1^o +
\upsilon \, t)$ and $\Omega_2(t) \equiv \Omega(z_2^o + \upsilon \,
t)$, and where the waist of the atom-laser coupling $\widetilde{w}$
is assumed to be much larger than for the cavity mode. With the
above couplings of the atoms to both, the laser and the cavity mode,
the atomic composite state can be manipulated in order to create an
energy exchange between $\ket{a, \bar{1}}$ and $\ket{1, \bar{a}}$ in
a similar way as have been suggested by You and coworkers. This
exchange is based on the sequence of four steps
\begin{equation}
   \ket{a, \bar{1};  n} \rightarrow \ket{a, \bar{e};  n}
   \begin{array}{c}
     \nearrow \raisebox{0.1cm}{$\ket{a, \bar{a}; \, n+1}$} \searrow \\
     \searrow \raisebox{-0.1cm}{$\ket{e, \bar{e}; \, n-1}$} \nearrow
   \end{array}
   \ket{e, \bar{a};  n} \rightarrow \ket{1, \bar{a};  n}, \nonumber
\end{equation}
if there were $n$ photons initially in the cavity mode. As seen
above, this sequence contains in its middle part a virtual process
in which a photon is emitted by the first and absorbed by the second
atom, so that the final state of the atoms is independent of the
number of cavity photons. For an initially empty cavity, we can
therefore simplify the above sequence to
\begin{equation}
\label{process}
   \ket{a, \bar{1}; \, 0}
   \rightarrow \ket{a, \bar{e}; \, 0}
   \rightarrow \ket{a, \bar{a}; \, 1}
   \rightarrow \ket{e, \bar{a}; \, 0}
   \rightarrow \ket{1, \bar{a}; \, 0} \, .
\end{equation}

Therefore, if we omit to display the intermediate states in the
sequence (\ref{process}), the laser and cavity field together
produce an \textit{effective} atom-atom interaction evolution
$\ket{a, \bar{1}} \underset{L,C}{{\longrightarrow}} \ket{1,\bar{a}}$
in which the state of the cavity field is factorized out in the
vacuum state. By exploiting this effective evolution, the maximally
entangled state
\begin{equation}
\label{state0}
   \ket{\Phi} = \frac{1}{\sqrt{2}} \,
   \left( \ket{a, \bar{1}} + e^{i \varphi} \ket{1, \bar{a}} \right),
\end{equation}
where $e^{i \varphi}$ is a constant phase factor, can be generated
by tuning the atomic velocity $\upsilon$ and the inter-atomic
distance $\ell$ for a given set of cavity-laser parameters:
$\omega_C$, $\omega_L$, $w$, $g_o$, and $\Omega_o$. In the next
subsection, we shall analyze in more details how this effective
atom-atom interaction depends on the velocity and distance of the
atoms, while both atoms are passing through the setup.

After both the atoms $A_1$ and $A_2$ have left the cavity, the
electronic population of the excited states $\ket{a}$ and
$\ket{\bar{a}}$ is coherently transferred back to the (ground)
hyperfine levels $\ket{0}$ and $\ket{\bar{0}}$ in order to protect
them from the spontaneous decay of these levels. As before, this is
achieved by applying a Raman pulse $L_2$ behind the cavity [see
Fig.~\ref{fig:1}(c)]. The entangled state (\ref{state0}) is then
mapped onto the state
\begin{equation}
\label{state1}
   \ket{\Phi^\prime} = \frac{1}{\sqrt{2}} \left( \ket{0, \bar{1}} +
   e^{i \varphi} \ket{1, \bar{0}} \right) \, .
\end{equation}
All the manipulations with the atoms which we have just described
are summarized graphically in Fig.~\ref{fig:2}, in which the
spatio-temporal evolution of the atoms and the cavity is displayed.

\subsection{Time-Evolution of the Effective Atom-Atom Interaction}

While the sequence (\ref{process}) provides the basic idea of how an
effective coupling can be achieved between the atoms, we need to
analyze this sequence in more details as to understand how to
control this coupling in practice. For this purpose, we shall use
the \textit{adiabatic elimination} procedure (see Refs.~\cite{prl82,
prl85} and Ref.~\cite{pra75a} for another derivation) which enables
one to exclude all the intermediate degrees of freedom due to the
action of the cavity mode and the laser field.

Formally, the time evolution of the coupled atom-cavity-laser system
is driven by the Hamiltonian
\begin{equation}\label{ham0}
H = H_1 + H_2 + H_C,
\end{equation}
where ($\hbar = 1$, $\mu = 1,2$)
\begin{eqnarray}
\label{atomic-H} H_\mu & = &
      \omega_1 \ket{1}_\mu \bra{1} +
      \omega_e \ket{e}_\mu \bra{e} +
      \omega_a \ket{a}_\mu \bra{a} \nonumber
      \\
      & + & \frac{1}{2} \left[
         \Omega_\mu(t) e^{-i \omega_L t} \ket{e}_\mu \bra{1} +
         g_\mu(t) \, c \, \ket{e}_\mu \bra{a} + h.c.
      \right]; \nonumber
\end{eqnarray}
describes the atom $A_\mu$ and its interaction with the cavity and
laser field, and where
\begin{equation}
H_C  = \omega_C \, c^+ \, c, \nonumber
\end{equation}
refers to the cavity mode energy. In the atomic Hamiltonian
(\ref{ham0}), hereby $\hbar \omega_1$, $\hbar \omega_e$, and $\hbar
\omega_a$ are the (excitation) energies of atomic states $\ket{1}$,
$\ket{e}$ and $\ket{a}$ [see Fig.~\ref{fig:1}(a)], while $c$ and
$c^+$ denote the annihilation and creation operators for a photon in
the cavity mode which act upon the Fock states $\ket{n}$.

In order to simplify the evaluation of the Schr\"{o}dinger equation
that is associated with the Hamiltonian (\ref{ham0}), let us switch
here to the interaction picture given by \cite{pra67}
\begin{eqnarray}\label{picture}
U^0_{int} & = &
     e^{-\im (\omega_1 + \omega_L) \, t \sum_\mu \ket{e}_\mu \bra{e} -
     \im \omega_1 t \sum_\mu \ket{1}_\mu \bra{1}} \; \times \nonumber
     \\
     && e^{-\im \omega_a t \sum_\mu \ket{a}_\mu \bra{a} -
     \im [ \omega_L - (\omega_a - \omega_1)] \, t \, c^+ c} \, .
\end{eqnarray}
In this picture, the atom-cavity-laser interaction Hamiltonian
becomes
\begin{eqnarray}\label{ham1}
H_{int} & = &
        - \delta \, c^+ c + \Delta \sum_\mu \ket{e}_\mu \bra{e}
        \\
        & + & \frac{1}{2} \sum_\mu \left[
         \Omega_\mu(t) \ket{e}_\mu \bra{1} +
         g_\mu(t) \, c \, \ket{e}_\mu \bra{a} + h.c.
        \right] \, , \nonumber
\end{eqnarray}
where $\Delta = \omega_{e1} - \omega_L$ and $\delta = \omega_L -
\omega_C - \omega_{a1} = (\omega_{ea} - \omega_C) - (\omega_{e1} -
\omega_L)$ refer to the off-resonance shifts (detuning) of the laser
and cavity frequencies as depicted in Fig.~\ref{fig:1}(b).

The Hamiltonian (\ref{ham1}) drives the state of the composite
atom-cavity-laser system due to the Schr\"{o}dinger equation
\begin{equation}\label{eq1}
\im \frac{d \ket{\Psi(t)}}{dt} = H_{int} \ket{\Psi(t)},
\end{equation}
where the (composite) wave function $\ket{\Psi(t)}$ is defined in
the product space of three (sub)systems: $A_1 (\ket{1}, \ket{e},
\ket{a})$, $A_2 (\ket{\bar{1}}, \ket{\bar{e}}, \ket{\bar{a}})$, and
the cavity Fock states $C (\ket{0}, \ket{1})$. Moreover, by taking
into account the composite states that occur in sequence
(\ref{process}), we may restrict this wave function to the subspace
\begin{eqnarray}\label{wf1}
\ket{\Psi(t)} & = &
               C_1(t) \ket{a, \bar{1}; \, 0} +
               C_2(t) \ket{a, \bar{e}; \, 0} +
               C_3(t) \ket{a, \bar{a}; \, 1} \nonumber
              \\[0.2cm]
              && + \,
                C_4(t) \ket{e, \bar{a}; \, 0} +
                C_5(t) \ket{1, \bar{a}; \, 0},
\end{eqnarray}
for which the Schr\"{o}dinger equation (\ref{eq1}) gives rise to the
set of closed equations
\numparts
\begin{eqnarray}\label{eq2}
   \im \dot{C}_1(t) &=& \frac{1}{2} \Omega_2(t) \, C_2(t), \label{eq2a}
   \\[0.1cm]
   \im \dot{C}_2(t) &=& \Delta \, C_2(t) + g_2(t) \, C_3(t) +
   \frac{1}{2} \Omega_2(t) \, C_1(t), \label{eq2b}
   \\[0.1cm]
   \im \dot{C}_3(t) &=& -\delta \, C_3(t) + g_1(t) \, C_4(t) +  g_2(t)
   \, C_2(t), \label{eq2c}
   \\[0.1cm]
   \im \dot{C}_4(t) &=& \Delta \, C_4(t) + g_1(t) \, C_3(t) +
   \frac{1}{2} \Omega_1(t) \, C_5(t), \label{eq2d}
   \\[0.1cm]
   \im \dot{C}_5(t) &=& \frac{1}{2} \Omega_1(t) \, C_4(t), \label{eq2e}
\end{eqnarray}
\endnumparts
and where the dot denotes the time derivative.

The off-resonant regime of the atom-cavity and atom-laser
interactions, we assumed, implies
\begin{equation}\label{cond}
   |\delta| \gg |g_\mu(t)|, \quad
   |\Delta| \gg |\Omega_\mu(t)|, \quad
   |\delta \, \Delta| \gg \left| g_{\mu}^2(t) \right| \, .
\end{equation}
These conditions, therefore, justify the adiabatic elimination
procedure for a sufficiently slow-varying time-dependent atom-cavity
$g_\mu(t)$ and atom-laser coupling $\Omega_\mu(t)$. The adiabatic
elimination procedure implies the vanishing of the time derivatives
$\dot{C}_2(t)$, $\dot{C}_3(t)$, and $\dot{C}_4(t)$, which together
with the conditions (\ref{cond}) lead to the exclusion of the
Eqs.~(\ref{eq2b})-(\ref{eq2d}) that account for the evolution of the
state vectors $\ket{a, \bar{e}; \, 0}$, $\ket{a, \bar{a}; \, 1}$,
and $\ket{e, \bar{a}; \, 0}$, respectively. Here, we shall omit the
details of the derivation for which the reader is refereed to the
literature \cite{pra67, prl82, prl85, pra75a}. The remaining
Eq.~(\ref{eq2a}) and Eq.~(\ref{eq2e}) for the functions $C_1(t)$ and
$C_2(t)$ take the closed form
\numparts
\begin{eqnarray}
   \im \dot{C}_1(t) & = & - \frac{\Omega_{2}^2(t)}{4 \, \Delta} \, C_1(t)
   + \lambda(t) \, C_5(t), \label{eq3a}
   \\
   \im \dot{C}_5(t) & = & \lambda(t) \, C_1(t) -
   \frac{\Omega_{1}^2(t)}{4 \, \Delta} \, C_5(t), \label{eq3b}
\end{eqnarray}
\endnumparts
where
\begin{equation}
\lambda(t) = \frac{\Omega_1(t) \, \Omega_2(t) \, g_1(t) \, g_2(t)}{4
\, \delta \, \Delta^2} \nonumber
\end{equation}
is the effective coupling between the initial and final composite
states $\ket{a, \bar{1}; \, 0}$ and $\ket{1, \bar{a}; \,0}$,
respectively.

The atom-laser coupling $\Omega_{\mu}(t)$ is determined by the
interaction of the electric-dipole of the atom with the electric
field of the laser. However, since the waist of the laser beam is
assumed to be much larger than those of the cavity mode, we may take
$\Omega_{\mu}(t) = \Omega = const.$ and include the time-variation
only due to the atom-cavity coupling $g_{\mu}(t)$. With this
simplification in mind, an analytical solution of
Eqs.~(\ref{eq3a})-(\ref{eq3b}) can be obtained in the form
\begin{eqnarray}
\label{wf3}
   \ket{\Psi(t)} & = & e^{\im \frac{\Omega^2}{4 \Delta} t}
   \ket{\Phi(t)}
\end{eqnarray}
with
\begin{eqnarray}
\label{wf2}
   \ket{\Phi(t)} & = & \cos \xi(t) \ket{a, \bar{1}} - \im \sin \xi(t)
   \ket{1, \bar{a}} \, ,
\end{eqnarray}
if the wave-function $\ket{\Psi(t)}$ was prepared initially in the
product state $\ket{a, \bar{1}; \, 0}$. In the expression
(\ref{wf2}), moreover, the cavity field state is not shown as being
factorized out in the vacuum state and the effective atom-atom
\textit{coupling angle} is given by
\begin{equation}
   \xi(t) = \int_{- \infty}^t \lambda(s) \, ds. \nonumber
\end{equation}

The wave function (\ref{wf2}) describes an entangled state for the
atoms $A_1$ and $A_2$, whose time evolution can be obtained also
from the effective Hamiltonian
\begin{equation}\label{ham2}
   H_{eff} = \lambda(t) \left( \sigma_1^{-} \sigma_2^{+} +
                                      \sigma_1^{+} \sigma_2^{-} \right) \, ,
\end{equation}
where $\sigma^+_\mu = \ket{1}_\mu \bra{a}$ and $\sigma^-_\mu =
\ket{a}_\mu \bra{1}$ denote the two-photon atomic excitation and
de-excitation operators. Owing to its obvious simplicity, this
Hamiltonian provides a much better understanding of the effective
two-atom evolution (\ref{wf2}) that is mediated by the cavity-laser
fields and by using the ansatz (\ref{wf1}) within the adiabatic
regime. Below, we shall restrict ourselves to the evolution of the
function $ \ket{\Phi(t)}$ since the Hamiltonian that drives the wave
function $\ket{\Psi(t)}$ differs from (\ref{ham2}) by just the
constant term $H_0 = \frac{\Omega^2}{4 \Delta} \left( \ket{1}\bra{1}
+ \ket{\bar{1}}\bra{\bar{1}} \right)$. This factor need not to be
considered since we could utilize a modified interaction picture
given by the unitary transformation $U^1_{int} = \exp \left( \im H_0
t \right)$, for which the (original) wave function (\ref{wf3}) would
coincide with (\ref{wf2}).

When both atoms have left the cavity (which is formally obtained in
the limit $t \rightarrow +\infty$), the state (\ref{wf2}) becomes
\begin{equation}\label{eq5}
   \ket{\Phi_{+\infty}} = \cos \theta(\upsilon, \ell) \: \ket{a, \bar{1}}
   - \im \sin \theta(\upsilon, \ell) \: \ket{1, \bar{a}} \, ,
\end{equation}
and where the asymptotic coupling angle is given by
\begin{equation}\label{theta}
   \theta(\upsilon, \ell) \equiv \xi(+\infty) = \sqrt{\frac{\pi}{32}}
   \, \frac{\Omega^2 \, g^2_o \, w}{\delta \, \Delta^2 \, \upsilon} \,
   \exp \left( -\frac{\ell^2}{2 \, w^2} \right) \, .
\end{equation}
Note that according to our scheme in Fig.~\ref{fig:2}, the atomic
states $\ket{a}$ and $\ket{\bar{a}}$ are mapped onto the hyperfine
states $\ket{0}$ and $\ket{\bar{0}}$ by applying a Raman pulse $L_2$
shortly after the atoms have crossed the cavity. Therefore, the
wave-function (\ref{eq5}) becomes
\begin{equation}\label{eq6}
   \ket{\Phi^\prime_{+\infty}} = \cos \theta(\upsilon, \ell) \ket{0,
   \bar{1}} - \im \sin \theta(\upsilon, \ell) \ket{1, \bar{0}}.
\end{equation}
\begin{figure}[t]
\begin{center}
\includegraphics[width=0.95\textwidth]{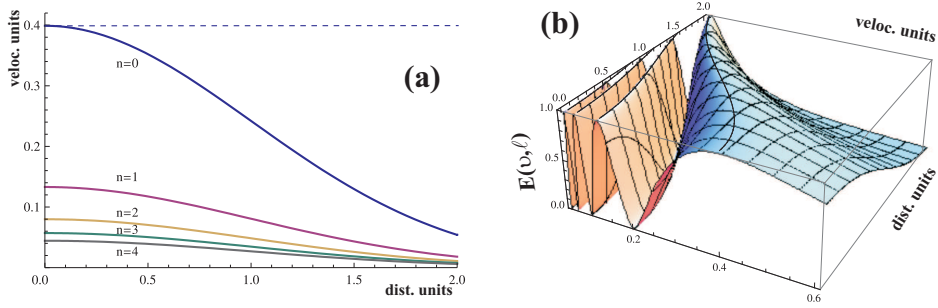}
\\[0.5cm]
\caption{(Color online) (a) Atomic velocities $\upsilon$ and
inter-atomic distances $\ell$ for which the initial product state
$\ket{a, \bar{1}}$ becomes maximally entangled due the cavity-laser
mediated atom-atom interaction. The velocity $\upsilon$ is displayed
in units of $\Omega^2 \, g^2_o \, w / \delta \, \Delta^2$ and the
inter-atomic distance in units of $w$. Along the lines, the
condition $\theta(\upsilon, \ell) = (2n + 1) \, \pi/4$ is satisfied
for the asymptotic couplings angle with $n = 0,1,2,3,4$. The
straight dashed line corresponds to a vanishing inter-atomic
distance, which is obtained in the formal limit $\ell \rightarrow 0$
for $n = 0$. (b) Von Neumann entropy $E(\upsilon, \ell)$ as function
of the atomic velocity $\upsilon$ and inter-atomic distance $\ell$
(using the same units).}
\label{fig:3}
\end{center}
\end{figure}

From Eq.~(\ref{eq6}), we can easily read off the condition:
$\theta(\upsilon, \ell) = (2n + 1) \, \pi/4$, with $n$ being an
integer, for which the two atoms become maximally entangled with
each other initially being prepared in the product state $\ket{a,
\bar{1}}$. For fixed cavity-laser parameters ($\delta$, $\Delta$,
$w$, $g_o$ and $\Omega$), this condition implies that the values of
atomic velocity $\upsilon$ and inter-atomic distances $\ell$ cannot
be chosen arbitrarily but must follow the (so-called) \textit{lines
of maximal entanglement} displayed in Fig.~\ref{fig:3}(a) for $n =
0,1,2,3,4$. According to this figure, the change between the
(maximally) entangled and disentangled state occurs more and more
rapidly as the velocity is decreased (or $n$ increases). In
Fig.~\ref{fig:3}(a), all velocities are given in units of $\Omega^2
\, g^2_o \, w / \delta \, \Delta^2$ and all distances in units of
the cavity waist $w$. For typical atom-cavity-laser parameters:
$\delta = 360$ MHz, $\Delta = 380$ MHz, $g_o = 27$ MHz, $\Omega =
50$ MHz, and $w = 13 \, \mu$m, these velocity and distance units
take the values of 0.46 m/s and 13 $\mu$m, respectively. These
values are compatible with the velocities in the range $0.01,
\ldots, 1$ m/s which were utilized in the recent cavity QED
experiments \cite{prl98, prl95a, njp10}, in which atoms are
coherently transported inside the cavity by means of a optical
lattice trap (see below).

Next, let us analyze how the degree of entanglement depends on the
velocity $\upsilon$ and distance $\ell$ of the atoms. For this
reason we display in Fig.~\ref{fig:3}(b) the \textit{von Neumann
entropy} \cite{nc}
\begin{eqnarray}\label{entropy}
   E(\upsilon, \ell)
   & \equiv &
   - \mbox{Tr} \left[ \rho (\upsilon,
   \ell) \, \log_2 \rho (\upsilon, \ell) \right] \nonumber
   \\
   & = &
   -\cos^2 \theta(\upsilon, \ell) \, \log_2 [\cos^2 \theta(\upsilon,
   \ell)] \nonumber
   \\
   && -\sin^2 \theta(\upsilon, \ell) \, \log_2
   [\sin^2 \theta(\upsilon, \ell)] \, ,
\end{eqnarray}
where $\rho (\upsilon, \ell) = \mbox{Tr}_2
(\ket{\Phi^\prime_{+\infty}}
 \bra{\Phi^\prime_{+\infty}})$ denotes the reduced density operator of
the first hyperfine qubit [see Eq.~(\ref{eq6})]. The maximal values
of the von Neumann entropy, i.e., $E(\upsilon, \ell) = 1$, are
obtained for the velocities and distances as displayed in
Fig.~\ref{fig:3}(a). Moreover, as seen from Fig.~\ref{fig:3}(b), the
velocities and distances along the (blue) $n=0$ line from
Fig.~\ref{fig:3}(a) appear to be the most appropriate for any
practical implementation of this scheme, since for these values of
$\upsilon$ and $\ell$, the obtained entanglement is less sensitive
with regard to small uncertainties. This leads us to the conclusion
that the $\upsilon$ and $\ell$ combinations along this line ($n=0$)
might be relevant for experimental attempts to generate the
atom-atom entanglement by means of the suggested setup.

Since the atom-atom interaction sequence (\ref{process}) can be
easily time reversed to
\begin{equation}\label{process2}
   \ket{1, \bar{a}; \, 0} \rightarrow \ket{e, \bar{a}; \, 0}
   \rightarrow \ket{a, \bar{a}; \, 1} \rightarrow \ket{a, \bar{e}; \,0}
   \rightarrow \ket{a, \bar{1}; \, 0}  \nonumber
\end{equation}
we can generate also the state ($t \rightarrow +\infty$)
\begin{equation}\label{eq7}
   \ket{\widetilde{\Phi}_{+\infty}} = \cos \theta(\upsilon, \ell) \:
   \ket{1, \bar{a}} - \im \sin \theta(\upsilon, \ell) \: \ket{a, \bar{1}}, \,
\end{equation}
from the atoms initially being prepared in the product state
$\ket{1, \bar{a}}$. Together with the Raman pulse $L_2$ that maps
back the atomic states $\ket{a} \rightarrow \ket{0}$ and
$\ket{\bar{a}} \rightarrow \ket{\bar{0}}$, we then obtain the state
\begin{equation}\label{eq8}
   \ket{\widetilde{\Phi}^\prime_{+\infty}} =
   \cos \theta(\upsilon, \ell) \ket{1, \bar{0}}
   - \im \sin \theta(\upsilon, \ell) \ket{0, \bar{1}} \, .
\end{equation}
For the other two initial (product) states $\ket{a, \bar{a}}$ and
$\ket{1, \bar{1}}$, in contrast, no effective interaction occurs on
the atoms when they pass through the cavity-laser system. From this
fact and Eqs.~(\ref{eq5}), (\ref{eq7}), we conclude that the
effective Hamiltonian (\ref{ham2}) gives a complete description of
the (effective) atom-atom interaction for all four possible initial
product states of the two atoms being mediated by the cavity-laser
fields in the adiabatic regime.

\begin{figure}
\begin{center}
\includegraphics[width=0.95\textwidth]{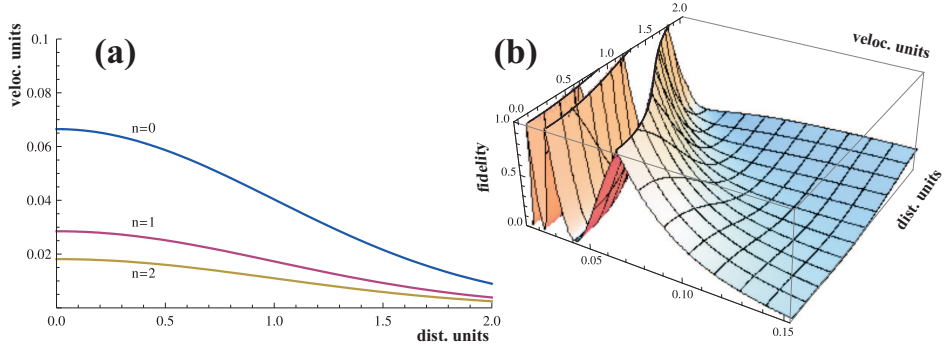}
\\[0.1cm]
\caption{(Color online) (a) Combinations of the atomic velocity
$\upsilon$ and inter-atomic distance $\ell$ that realizes the i-swap
gate (\ref{iswap}), i.e., for which the condition (\ref{cond1}) is
fulfilled for $n = 0,1,2$. (b) Fidelity $F_{i-swap}(\upsilon, \ell)$
as function of the atomic velocity $\upsilon$ and inter-atomic
distance $\ell$. The same units of $\upsilon$ and $\ell$ are used as
in Fig.~\ref{fig:3}.}
\label{fig:4}
\end{center}
\end{figure}

\section{Two-Qubit Quantum Logic Gates}

In the previous section, we have shown how the atomic hyperfine
qubits of the two atoms $A_1$ and $A_2$ can be manipulated
adiabatically by means of the cavity-laser setup from
Fig.~\ref{fig:1}(c) and the sequence of steps from Fig.~\ref{fig:2}.
Independent of the initial state of the qubits, the evolution of the
two-qubit hyperfine input state $\ket{\psi_{in}} = \sum_i c_i^o
\ket{\mathbf{v}_i}$ into the output state $\ket{\psi_{out}} = \sum_i
c_i (\upsilon, \ell) \ket{\mathbf{v}_i}$ ($i,j = 1,\ldots,4$) is
given by the unitary matrix
\begin{equation}\label{matrix}
   \hspace*{-0.2cm}
   U_{ij}(\upsilon, \ell) = \left(
   \begin{array}{cccc}
   1 & 0 & 0 & 0 \\
   0 & \cos \theta(\upsilon, \ell) & - \im \sin \theta(\upsilon, \ell) & 0 \\
   0 & - \im \sin \theta(\upsilon, \ell) & \cos \theta(\upsilon, \ell) & 0 \\
   0 & 0 & 0 & 1
   \end{array} \right)
\end{equation}
expressed in the two-qubit hyperfine basis
\begin{equation}\label{basis}
   \ket{\mathbf{v}_1} = \ket{0, \bar{0}}, \,
   \ket{\mathbf{v}_2} = \ket{0, \bar{1}},
   \ket{\mathbf{v}_3} = \ket{1, \bar{0}}, \,
   \ket{\mathbf{v}_4} = \ket{1, \bar{1}} \, ,
\end{equation}
and where $c_i (\upsilon, \ell) = \sum_j U_{ij}(\upsilon, \ell) \,
c_j^o$. For different values of the atomic velocity $\upsilon$ and
inter-atomic distance $\ell$, different transformations are
therefore realized including, for instance, the generation of
maximally entangled state (\ref{state1}) if one starts from the
initial product state $\ket{0, \bar{1}}$. Moreover, we can analyze
the atom-atom coupling angle $\theta(\upsilon, \ell)$ for different
combinations of $\upsilon$ and $\ell$ and for its capabilitiy to
realize non-trivial two-qubit quantum gates. In fact, the suggested
set-up is suitable for realizing the i-swap, controlled-Z, and the
controlled-$\overline{\mbox{NOT}}$ quantum gates for different
choices of the velocity and distance, together with some minor
modifications in the steps that are necessary to prepare the atoms
before (afterwards) they enter (leave) the cavity-laser system (see
below). In the following, we consider these gates in more details
and display their temporal diagrams and possible values $(\upsilon,
\ell)$ for which these gates are realized.

\subsection{i-Swap Gate}

Perhaps the simplest quantum gate is the i-swap gate \cite{pra67_1}
which is expressed in the atomic basis (\ref{basis}) as
\begin{equation}\label{iswap}
   U_{ij}^{i-swap} = \left(
   \begin{array}{cccc}
   1 & 0 & 0 & 0 \\
   0 & 0 & \im & 0 \\
   0 & \im & 0 & 0 \\
   0 & 0 & 0 & 1
   \end{array} \right) \, .
\end{equation}
By comparing the Eqs.\ (\ref{matrix}) and (\ref{iswap}), we see that
this gate can generated whenever the effective coupling angle
fulfills the condition
\begin{equation}\label{cond1}
   \theta(\upsilon, \ell) = 3\pi / 2 + 2 \pi n \, .
\end{equation}
The Fig.~\ref{fig:4}(a) displays combinations of the atomic velocity
$\upsilon$ and the inter-atomic distance $\ell$ which satisfy this
condition for $n=0,1,2$. For the i-swap gate, moreover, the sequence
of steps that needs to be carried out before and after the atoms
have crossed the cavity is the same as shown in Figure~\ref{fig:2}
and no additional manipulations are required in order to implement
this gate.

From the viewpoint of an experiment, as we have discussed, it is
important to know how stable a gate operation can be performed for
small deviations in the $(\upsilon,\ell)$ parameters. This stability
can be seen from the fidelity (distance) between the i-swap gate
(\ref{iswap}) and the unitary matrix (\ref{matrix}) obtained for
different values of $(\upsilon,\ell)$. The Fig.~\ref{fig:4}(b)
displays such a fidelity that we have defined as
\begin{eqnarray}\label{fidelity1}
   F_{i-swap}(\upsilon, \ell)
   & \equiv & 1 - \mathcal{N} \left( \| U(\upsilon, \ell)
                - U^{i-swap} \| \right) \nonumber \\
   & = & 1 -  \sqrt{\frac{1 + \sin \theta(\upsilon, \ell)}{2}},
\end{eqnarray}
where $\| M \| \equiv \sqrt{\mbox{Tr} \left( M \, M^+ \right)}$ is
the \textit{Frobenius} norm \cite{gvl} and $\mathcal{N}\left(
f_{\upsilon, \ell} \right) \equiv f_{\upsilon, \ell} \cdot \left(
\mbox{Max}(f_{\upsilon, \ell} \right))^{-1}$ is used for its
normalization upon the interval $0 \leq F_{i-swap} \leq 1$.

By construction, this fidelity is a continuous function for which
the realization of the i-swap gate occurs when $F_{i-swap}
(\upsilon, \ell) = 1$, which corresponds to the values $( \upsilon,
\ell )$ displayed in Fig.~\ref{fig:4}(a).

\begin{figure}
\begin{center}
\includegraphics[width=0.95\textwidth]{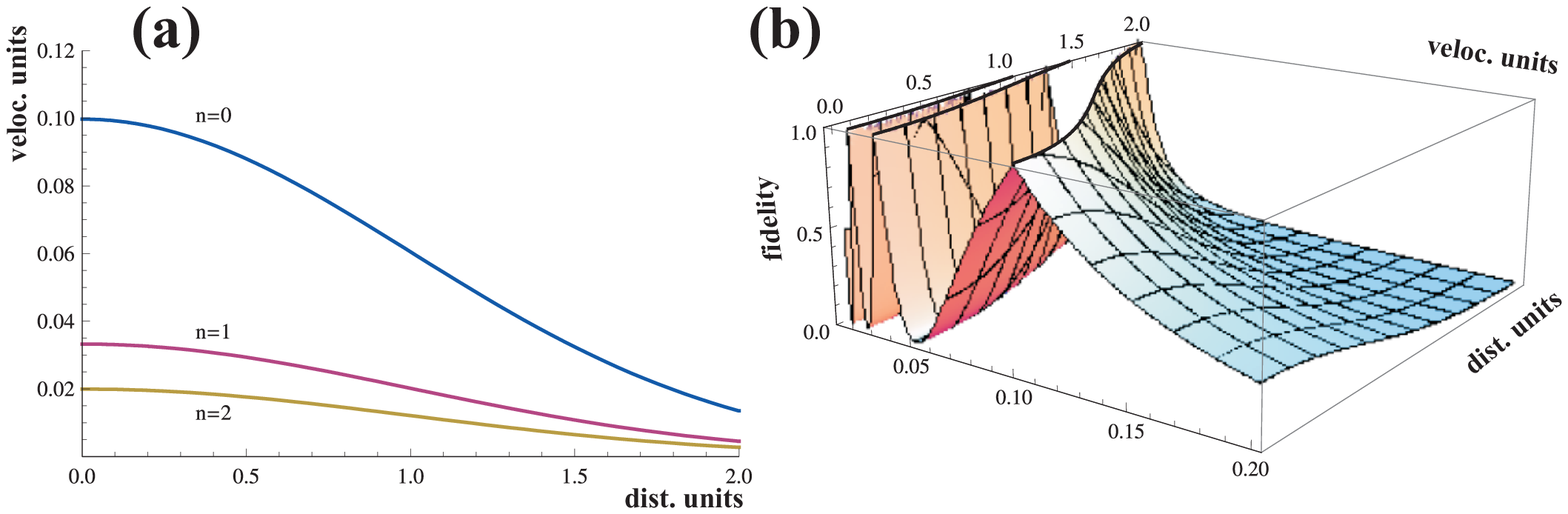}
\\[0.5cm]
\includegraphics[width=0.95\textwidth]{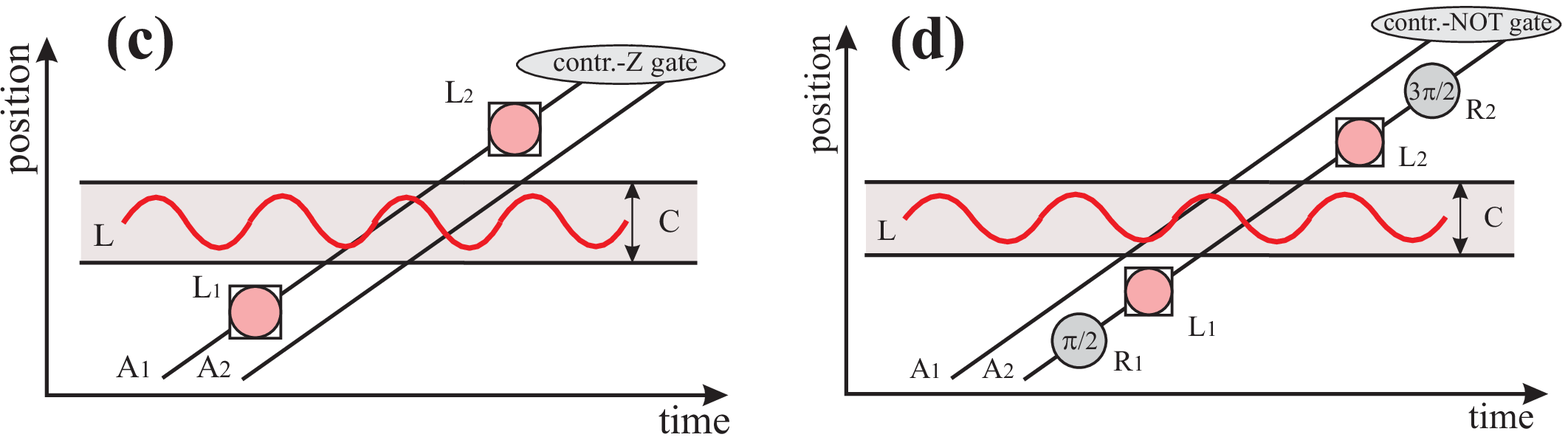}
\\[0.1cm]
\caption{(Color online) (a) Combinations of the atomic velocity
$\upsilon$ and inter-atomic distance $\ell$ that realizes the
controlled-Z (\ref{qpg}) and the controlled-$\overline{\mbox{NOT}}$
(\ref{cnot}) gates, i.e., that satisfy the conditions (\ref{cond2})
and (\ref{cond3}) for $n = 0,1,2$. (b) Fidelity $F_{CZ}(\upsilon,
\ell)$ as function of the atomic velocity $\upsilon$ and
inter-atomic distance $\ell$. The same units of $\upsilon$ and
$\ell$ are used as in Fig.~\ref{fig:3}. (c) Temporal diagram for
generating the controlled-Z gate for the two hyperfine qubits of the
atoms $A_1$ and $A_2$. (d) The same as in Fig.~\ref{fig:5}(c) but
for the controlled-$\overline{\mbox{NOT}}$ gate.}
\label{fig:5}
\end{center}
\end{figure}

\subsection{Controlled-Z Gate}

For two interacting qubits $A$ and $B$, the controlled-Z gate is
defined by the transformation \cite{nc}
\begin{equation}\label{qpg-def}
   U_{CZ} \ket{\alpha_A, \beta_B} =
   (-1)^{\alpha \cdot \beta} \ket{\alpha_A, \beta_B},
\end{equation}
where $\alpha, \beta= 0,1$ are the basis states. This gate is a
simple example of the conditional quantum dynamics which introduces
an additional phase $e^{i \pi} = -1$ whenever both qubits are in the
state $\ket{1_A, 1_B}$.

In Sec.~2, we concluded that the initial product state $\ket{1,
\bar{1}}$ of the two atoms does not undergo any evolution mediated
by the cavity-laser fields. Therefore, the direct identification of
the atomic hyperfine states $\ket{0},\, \ket{1}$ and
$\ket{\bar{0}},\, \ket{\bar{1}}$ with the (logical) qubit states
$\ket{0_A},\, \ket{1_A}$ and $\ket{\bar{0}_B},\, \ket{\bar{1}_B}$ in
(\ref{qpg-def}) will not allow us to realize the control-Z gate,
while the reversed assignment for the qubit $A$
\begin{equation}
\ket{0} = \ket{1_A}, \quad \ket{1} = \ket{0_A}, \quad \ket{\bar{0}}
= \ket{0_B}, \quad \ket{\bar{1}} = \ket{1_B}. \nonumber
\end{equation}
would do so. With this assignment of the basis (\ref{basis}), the
transformation matrix for the requested controlled-Z gate becomes
\begin{equation}\label{qpg}
U_{ij}^{CZ} = \left(
\begin{array}{cccc}
1 & 0 & 0 & 0 \\
0 & -1 & 0 & 0 \\
0 & 0 & 1 & 0 \\
0 & 0 & 0 & 1
\end{array} \right) \, .
\end{equation}

In contrast to the i-swap gate (\ref{iswap}), however, the matrix
(\ref{qpg}) cannot be obtained from the evolutionary matrix
(\ref{matrix}) by just imposing a condition of the type
(\ref{cond1}) on the coupling angle $\theta(\upsilon, \ell)$.
Instead we must consider here the new temporal diagram as displayed
in Fig.~\ref{fig:5}(c). The difference between this diagram and the
sequence from Fig.~\ref{fig:2} is that the second atom $A_2$ is not
subjected to the Raman pulses $L_1$ and $L_2$, implying that its
(hyperfine) state $\ket{\bar{0}}$ is not mapped upon the (optical)
state $\ket{\bar{a}}$ nor back. Due to suggested setup from
Fig.~\ref{fig:1}(c) this modification, for instance, is realized
simply by switching off the pairs of Raman laser beams while the
atom $A_2$ crosses the zones $L_1$ and $L_2$.

Following the temporal sequence in Fig.~\ref{fig:5}(c) and by making
use of Eq.~(\ref{eq5}), we see that the four input states will
evolve (after mapping $\ket{a} \rightarrow \ket{0}$) into
\begin{eqnarray}
\label{effective-qpg}
   \ket{0, \bar{0}} &\rightarrow& \ket{0, \bar{0}},     \nonumber \\
   \ket{0, \bar{1}} &\rightarrow&
   \cos \theta(\upsilon, \ell) \ket{0,\bar{1}}
   - \im \sin \theta(\upsilon, \ell) \ket{1, \bar{a}},            \\
   \ket{1, \bar{0}} &\rightarrow& \ket{1, \bar{0}},     \nonumber \\
   \ket{1, \bar{1}} &\rightarrow& \ket{1, \bar{1}} \, . \nonumber
\end{eqnarray}
when both atoms passed through the setup. Although the output state
in the second line does not belong to the basis set (\ref{basis}),
the transformation matrix (\ref{qpg}) is obtained whenever the
condition
\begin{equation}\label{cond2}
\theta(\upsilon, \ell) = \pi + 2 \pi n,
\end{equation}
is fulfilled. In this case, the unwanted part $\ket{1, \bar{a}}$ in
the second line vanishes. The Fig.~\ref{fig:5}(a) displays the
values of $\upsilon$ and $\ell$ for which the condition
(\ref{cond2}) is satisfied. Moreover, by applying the fidelity we
introduced in Sec.~3.1, the fidelity between the ideal gate
(\ref{qpg}) and the effective transformations (\ref{effective-qpg})
takes the form
\begin{equation}\label{fidelity2}
F_{CZ}(\upsilon, \ell) = 1 -  \sqrt{\frac{1 + \cos \theta(\upsilon,
\ell)}{2}},
\end{equation}
and is displayed in Fig.~\ref{fig:5}(b) for different values of
$\upsilon$ and $\ell$. As for the i-swap gate (\ref{iswap}), the
least rapid change in the fidelity occurs along the $n=0$ lines and,
in particular, for small interatomic distances but moderate
velocities.

\subsection{Controlled-$\overline{\mbox{NOT}}$ Gate}

For two interacting qubits $A$ and $B$, the
controlled-$\overline{\mbox{NOT}}$ gate is defined by the
transformation \cite{mynote}
\begin{equation}\label{cnot-def}
   U_{CN} = \ket{0_A}\bra{0_A} \times I^B - \ket{1_A}\bra{1_A} \times
   U^B_{\mbox{not}}
\end{equation}
where $I^B = \ket{0_B} \bra{0_B} + \ket{1_B} \bra{1_B}$ is the
identity matrix and $U^B_{\mbox{not}} = \ket{0_B} \bra{1_B} +
\ket{1_B} \bra{0_B}$ denotes the single-qubit NOT gate associated
with the qubit $B$. As for the standard controlled-$\mbox{NOT}$
gate, we shall refer to the qubits $A$ and $B$ as the
\textit{control} and \textit{target} qubit, respectively. While the
control qubit does not change its state under the gate
(\ref{cnot-def}), the target qubit is swapped together with the
phase factor $e^{i \pi} = -1$ when the control qubit is set to
$\ket{1_A}$. In the basis (\ref{basis}), the
controlled-$\overline{\mbox{NOT}}$ gate is therefore given by the
matrix
\begin{equation}\label{cnot}
   U_{ij}^{CN} = \left(
   \begin{array}{cccc}
   1 & 0 & 0 & 0 \\
   0 & 1 & 0 & 0 \\
   0 & 0 & 0 & -1 \\
   0 & 0 & -1 & 0
   \end{array} \right),
\end{equation}
where we utilized the assignment
\begin{equation}
\ket{0} = \ket{0_A}, \quad \ket{1} = \ket{1_A}, \quad \ket{\bar{0}}
= \ket{0_B}, \quad \ket{\bar{1}} = \ket{1_B}. \nonumber
\end{equation}

Obviously, the matrix (\ref{cnot}) can not be obtained from the
evolutionary matrix (\ref{matrix}) by just imposing a single
restriction on the coupling angle $\theta(\upsilon, \ell)$. Hence we
consider the modified temporal diagram  displayed in
Fig.~\ref{fig:5}(d). According to this diagram, the control atom
$A_1$ is not subjected to the Raman pulses $L_1$ and $L_2$, however,
the target atom $A_2$ passes through two additional classical
microwave fields, in which it undergoes the (coherent) rotation of
atomic hyperfine states
\numparts
\begin{eqnarray}
&&
  \ket{\bar{0}} \rightarrow \cos \left( \eta/2 \right) \ket{\bar{0}} -
   \sin \left( \eta/2 \right) \ket{\bar{1}}, \label{ramsey1} \\
&&
  \ket{\bar{1}} \rightarrow \sin \left( \eta/2 \right) \ket{\bar{0}} +
   \cos \left( \eta/2 \right) \ket{\bar{1}}, \, \label{ramsey2}
\end{eqnarray}
\endnumparts
where the rotation angle $\eta$ is proportional to the microwave
pulse duration. In the literature, such an atom-field interaction is
often called a Ramsey pulse and is denoted in Fig.~\ref{fig:5}(d) by
grey circles. These circles contain the interaction time in units of
Ramsey rotations $\eta$, and the letters $R_1$ and $R_2$ are
associated with the Ramsey zones in front and behind the cavity [see
Fig.~\ref{fig:1}(c)].

According to Fig.~\ref{fig:5}(d) and
Eqs.~(\ref{ramsey1})-(\ref{ramsey2}), the state of $A_2$ is first
transformed
\begin{equation}\label{super}
   \ket{\bar{0}}   \overset{\pi/2}{\rightarrow}
   \frac{1}{\sqrt{2}} \left( \ket{\bar{0}} - \ket{\bar{1}} \right)
   \quad \mbox{or} \quad
   \ket{\bar{1}}   \overset{\pi/2}{\rightarrow}
   \frac{1}{\sqrt{2}} \left( \ket{\bar{0}} + \ket{\bar{1}} \right)
\end{equation}
by using a $\pi/2$ Ramsey pulse in the zone $R_1$. Before the atom
enters the cavity, the atomic hyperfine state $\ket{\bar{0}}$ is
mapped upon the optical state $\ket{\bar{a}}$ by means of the Raman
pulse $L_1$, which overall, this gives rise to the superposition
\begin{equation}
  \ket{\bar{0}}  \underset{L_1}{\overset{\pi/2}{\longrightarrow}}
  \frac{1}{\sqrt{2}} \left( \ket{\bar{a}} - \ket{\bar{1}} \right)
  \quad \mbox{or} \quad \ket{\bar{1}}
  \underset{L_1}{\overset{\pi/2}{\longrightarrow}}
  \frac{1}{\sqrt{2}} \left( \ket{\bar{a}} + \ket{\bar{1}} \right) \, .
  \nonumber
\end{equation}
Inside the cavity, as mentioned above, only the product state
$\ket{1, \bar{a}}$ of the two atoms evolves according to
Eq.~(\ref{eq7}). This makes the target qubit $A_2$ to remain
unchanged if the control qubit $A_1$ was set initially to $\ket{0}$.
If the control qubit was set to $\ket{1}$, then the effective
atom-atom evolution (\ref{eq7}) applies and gives rise to a swap of
the target qubit $A_2$ for a proper choice of the velocity
$\upsilon$ and the inter-atomic distance $\ell$. When both atoms
have passed through the cavity, the state $\ket{\bar{a}}$ is mapped
back to $\ket{\bar{0}}$ by the Raman pulse $L_2$ and, finally, the
atom $A_2$ is subjected to a $3\pi/2$ Ramsey pulse in the zone
$R_2$.

The mentioned Ramsey and Raman pulses together with the cavity and
laser field make, therefore, the four input states of the hyperfine
qubits to evolve (up to global phase factor)
\numparts
\begin{eqnarray}
   \ket{0, \bar{0}}  &\rightarrow&   \ket{0, \bar{0}}, \label{e-cnot1} \\
   \ket{0, \bar{1}}  &\rightarrow&   \ket{0, \bar{1}}, \\
   \ket{1, \bar{0}}  &\rightarrow&
   \frac{\left( 1 + \cos \theta(\upsilon, \ell) \right)}{2} \ket{1,\bar{0}} -
   \frac{\left( 1 - \cos \theta(\upsilon, \ell) \right)}{2} \ket{1,\bar{1}}
   \nonumber   \\
   &&  \hspace{0.5cm}
   + \im \sin \theta(\upsilon, \ell)
   \frac{\left( \ket{a, \bar{0}} - \ket{a, \bar{1}} \right)}{2},
   \\
   \ket{1, \bar{1}}   &\rightarrow&
   \frac{\left( 1 + \cos \theta(\upsilon, \ell) \right)}{2} \ket{1,\bar{1}} -
   \frac{\left( 1 - \cos \theta(\upsilon, \ell) \right)}{2} \ket{1,\bar{0}}
   \nonumber \\
   &&  \hspace{0.5cm} + \im \sin \theta(\upsilon, \ell)
   \frac{\left(\ket{a, \bar{0}} - \ket{a, \bar{1}} \right)}{2} \, . \label{e-cnot4}
\end{eqnarray}
\endnumparts

Although, again, the output states in the last two lines do not
belong entirely to the basis set (\ref{basis}), the matrix
(\ref{cnot}) can be realized if we impose the condition
\begin{equation}\label{cond3}
\theta(\upsilon, \ell) = \pi + 2 \pi n
\end{equation}
for the effective coupling angle. Since it is the same condition as
Eq.~(\ref{cond2}) for the controlled-Z gate, the combinations of
$\upsilon$ and $\ell$ that are appropriate for the
controlled-$\overline{\mbox{NOT}}$ gate are displayed already in
Fig.~\ref{fig:5}(a) for $n=0,1,2$. The same applies also to the
fidelity that is displayed in Fig.~\ref{fig:5}(b).

\section{Summary and Outlook}

In summary, a scheme is proposed to generate an entangled state
between the hyperfine qubits of two non-interacting four-level atoms
being separated by a macroscopical distance. An effective
interaction between the atoms is mediated by a detuned optical
cavity and a laser beam. The purpose of our work is to analyze how
the position-dependent coupling of each atom to the same cavity mode
and a laser beam affects the effective interaction among the atoms,
and whether it is possible to create a (maximally) entangled state
between the atoms. In particular, the analytical expressions for the
(asymptotic) coupling angle (\ref{theta}) and the evolutionary
matrix (\ref{matrix}) tell us explicitly how the degree of
entanglement depends on both, the atomic velocity and the (initial)
inter-atomic distance. For a position-dependent atom-cavity coupling
(\ref{coupling}), these expression have been derived for the first
time for a four-level scheme as described above. In
Fig.~\ref{fig:3}(a), for instance, we have shown that the
atom-cavity interaction with a position independent atom-cavity
coupling, which was suggested in the Ref.~\cite{pra67}, leads to the
set of constant atomic velocities for which the maximally entangled
state can be generated (see dashed line). Under more realistic
assumptions of position-dependent atom-cavity coupling, however,
these velocities are not constant but depend on the values of
inter-atomic distance according to expression (\ref{theta}). From
Fig.~\ref{fig:3}(b), moreover, it can be seen how sensitive the
entanglement depends on variations in these parameters, an important
requisite for any experimental realization. Finally, a few realistic
schemes are suggested to implement some basic two-qubit quantum
gates, such as i-swap gate, controlled-Z, and the
controlled-$\overline{\mbox{NOT}}$ gate in the framework of the
given cavity-laser setup. For all these schemes, we displayed the
atomic velocities and inter-atomic distances for which these gates
are realized, i.e., the gate fidelities become maximal.

Following the recent experiments \cite{prl89, prl95a, njp10} and the
theoretical works of Refs.~\cite{pra75a, pra70a, pra75b, pra77a},
the position-dependent effects on the effective atom-atom
interaction and entanglement formation mediated by a (detuned)
optical cavity, are acknowledged today as a notable step in
obtaining the control over the entanglement of atoms within the
framework of cavity QED. In particular, Li and coworkers
\cite{pra75a} suggested that the distance between the atoms is an
important parameter that can be utilized to control the
(position-dependent) atom-cavity coupling which implies also the
control over the atomic entanglement. Instead of using a two-level
atomic configuration, however, for the cavities in optical domain it
appears more suitable to consider a four-level $\Lambda$-type level
configuration in which the quantum information is stored in the
hyperfine levels of the atomic ground state. Such a configuration
appears to be essential for the recent experimental attempts
\cite{prl98, prl95a, njp10} in which atoms are transported
coherently inside the cavity by means of a optical lattice trap
(conveyor belt). For such a belt, the inter-atomic distance is given
by the wavelength of the (standing) optical lattice, while the
velocity of the atoms is set by a shift in the frequencies of the
counter-propagating laser beams.

A further extension of the effective atom-atom evolution as
described in Sec.~2, might be a chain of $N$ four-level atoms that
cross the experimental setup in Fig.~\ref{fig:1}(c) and interacts
simultaneously with the same cavity mode while passing through the
cavity. This extension would lead naturally to the generation of
various $N$-partite entangled states depending on the $(\upsilon,
\ell)$ regime and the succession of Raman and Ramsey zones (see for
instance Ref.~\cite{pra77}, where we discussed the formation of
genuine entangled states for a chain of $N$ bi-level atoms which
cross an analogous experimental set-up we considered in this paper).
Finally we remark that an realistic atom-cavity interaction
evolution should also include the decoherence effects, which have
been avoided in this paper so far. We note that in order to analyze
the time evolution of such quantum systems embedded into a reservoir
or under the external noise and to analyze different entanglement or
separability measures, including those in Eqs.~(\ref{entropy}) and
(\ref{fidelity1}), a quantum simulator has been developed recently
in our group \cite{cpc175} that can be utilized for such studies in
future.

\ack This work was supported by the DFG under the project No. FR
1251/13.


\begin{thebibliography}{33}
\bibitem{mynote}
 In contrast to the controlled-$\overline{\mbox{NOT}}$ gate we have defined
 in Eq.~(\ref{cnot-def}), the conventional controlled-NOT gate \cite{nc}:
 $U = \ket{0_A}\bra{0_A} \times I^B +
      \ket{1_A}\bra{1_A} \times U^B_{\mbox{not}}$
 implies only the flipping of the target qubit (without acumulating a phase)
 when the control qubit is set to $\ket{1_A}$.
\end{thebibliography}
\end{document}